\documentclass{elsart}

\usepackage{graphicx}% Include figure files

\begin{document}

\begin{frontmatter}



\title{Finite size scaling approach to dynamic storage allocation problem}
%\date{\today}
\author{Hamed Seyed-allaei}
\ead{allaei@sissa.it}
\address{International School for Advanced Studies (SISSA) and INFM, 4,Via Beirut, 34014 Trieste, Italy}

\begin{abstract}
It is demonstrated how dynamic storage allocation algorithms can be analyzed in terms of finite size
scaling. The method is illustrated in the three simple cases of the {\it
first-fit, next-fit} and {\it best-fit} algorithms, and the system works at full capacity.
The analysis is done from two different points of view - running speed and employed memory. In both cases,
and for all algorithms, it is shown that a simple scaling function exists and the relevant exponents are
calculated. The method can be applied on similar problems as well.
\end{abstract}


\end{frontmatter}

%\maketitle

\section{Introduction}

Computers are growing rapidly.
The first IBM PC had $64,000$ bytes of memory, $160 kB$ of storage and a microprocessor with speed of 4.77 MHz\cite{firstPC}.
Today, they have $10^9$ bytes of memory and about
hundred times more space on their hard disks.
They can process billions of instructions just in one second.
These numbers are just for a typical desktop PC, while relative values for super computers could be many times greater.
At a higher level, computers can be connected to one another
to share their information and resources and make a large amount of
data available for users on a network like Internet.

These numbers are large enough to encourage us to manipulate
them using the tools of statistical physics.
These tools can help us to study and to estimate the efficiency
and performance of algorithms and methods which are designed to organize
and deal with large amounts of data and resources.
Some examples of these algorithms are storage allocation algorithms and file systems and
search and indexing algorithms.

In this work we will present finite size scaling \cite{scaling} as a
tool to study dynamic storage allocation algorithms.
We are going to use the same terminology of critical phenomena in the
present context.

We have chosen a well defined problem \cite{knuth1} - a
comparison of speed and memory usage among three simple allocation algorithms:
first-fit, next-fit and best-fit algorithm.
Nevertheless, this can also be done with other algorithms which are
applied on a large set of data and can give us a better understanding
of what their behaviors are especially when there is not a simple analytical calculation.

First we should give a brief description of storage allocation methods.
A storage device such as memory or hard-disk-drive (HDD) is a pool of {\it words}
(we will call word the smallest addressable unit of memory) \cite{knuth1}
and an integer number, the address, shows the position of each word in the memory.

The goal of a memory allocation routine is to store and organize data in
memory in the most efficient manner - to allocate enough memory at any new
request, and to restrict access to the allocated memory from other processes until
the task requiring memory liberates it\cite{knuth1}.

Since the size of any request is a dynamical variable
which is out of storage allocator's
control, it is called dynamic storage allocation and it is a sort of on-line
algorithm \cite{wilson}.

One simple way to allocate memory consists of dividing memory into large and uniform
{\it nodes} (the smallest part of a data structure including at least one word)\cite{knuth1}.
Since data sizes can be smaller than the size of the nodes, this method will waste large amount of memory.
Alternatively one might divide memory pool into small nodes and use the scheme of the {\it
linked-memory}: ``if there is no room for the information in a given location, put
it elsewhere and implant a link to it'' \cite{knuth1}. Because many users and
applications may need different size blocks of memory and may need them for different periods of time, data will be
fragmented after some time. Data fragmentation slows down reading speed
especially when the storage device is a hard disk \cite{microsoft}.

It is feasible to allocate space for a given request just in one
contiguous {\it block} (a set of contiguous memory locations)\cite{knuth1}.
We can avoid data fragmentation in this fashion, but still have problem with free space fragmentation.
Indeed, after the dynamic storage allocation has been in operation for a while, the free areas are broken
into many small pieces which might not be large enough for certain requests;
therefore, a larger amount of memory is required
to run the system in this way.
This amount of memory depends on the two main parts of the algorithm - the method of keeping list of free blocks and
the selection policy for choosing among the available free blocks of memory.
The list of available free blocks can help in their better use. It stores the addresses and sizes of
free blocks.
Indeed there are many different types of free lists: address-ordered, size-ordered and
etc.\cite{wilson}. The simplest one which we will use, is the address-ordered list. It
keeps the free blocks with the order of their address in storage device\cite{wilson}.
in fact, the order of
available block in the list is the same as their place in the memory. For example if a free
$b_n$ starts from word number $n$ and another block $b_m$ starts from word number  $m$, in the
list of free blocks, block $b_n$ is before $b_m$, only when $n \le m$.

Given the list of free blocks, we need a decision policy which tells us where
a request should be placed. One of the simplest policies is known as {\it
first-fit} \cite{knuth1}: it tries to find the first available block in the free list that
is large enough for the request. The obvious disadvantage of this algorithm is that
it might waste a large space for other future requests by putting a small request there.
Another issue with the first-fit is that after the system has been working for a while, the
available list will be sorted by increasing order of blocks' size\cite{knuth1}.
Thus, for a large request, we have to check almost all of the list of free blocks
which slows down the running speed of the algorithm.

If it is desired to reduce memory waste, another method, {\it best-fit} algorithm\cite{knuth1}
can be applied. This method tries to find the smallest available space,
which is large enough for the request; therefore, we can reduce space use at the cost of more searching in the free blocks' list.

If speed of the algorithm is important for us we can use next-fit algorithm which
is a variation of first-fit algorithm. We define a pointer which
points to the starting point of the searching in the list. The value of the pointer will be changed after a successful allocation and
it will point to the first available block
after
%that can be the rest of the
the free block used for the last allocation.
In this way the size distribution of free blocks in the free list will be uniform
hence the algorithm will be faster\cite{knuth1}.

There are many other algorithms such as {\it buddy-system, next-fit} and {\it indexed-fit} algorithms\cite{knuth1,wilson}.
A particular choice might be dictated by bottlenecks
of the storage device, size distribution function of requests or their life span.
For example if a large memory is available the {\it buddy-system} may be used
because it is
fast but needs more space to run. In the opposite case when there is shortage of memory,
the {\it linked-memory} algorithm is more appropriate. However, in most cases there is not any clear
way to determine which algorithm is the best. Many attempts have been made; these attempts
answer this question and compare different algorithms from different points
of view like performance or probability to have the worst case fragmentation\cite{worst,nielsen}.

We aim to study and compare speed and memory usage, using tools of statistical physics like finite size
scaling for  simple algorithms: the first-fit, next-fit and best-fit algorithms with {\it
address-ordered} free list.

It is shown that best-fit algorithm is the best in terms of space use while the next-fit algorithm is the worst.
From the speed point of view, the results show that next-fit algorithm is the fastest and there is not
a significant difference between the speed of the other two.

\section{Model}
The model we will study assumes a storage device such as memory or hard disk of a
computer as a set of small nodes.
At the $r$-th step of simulation, a request of size
$S_{r}$ is sent to the allocator, $S_{r}$ being the number of nodes (or words) needed.
We have performed a detailed study of the case where the size $S_{r}$ is drawn from a
uniform distribution $1\le S_{r}\le S_{max}$. Although is straightforward to
extend the analysis to other interesting cases the uniform distribution has often been
used previously \cite{wilson}.
Furthermore at each step of the simulation each allocated block can be liberated with the
probability $P_d$. This is a sort of ``grand canonical ensemble'' where  the total number
of blocks, $N_{b}$, fluctuates around its average value, $1/P_d$.

In order to prevent the rejection of a request due to an overflow, we allow the memory to
increase by the necessary amount when needed. In fact in this way, the memory size is a
variable which can be measured during the simulation. This is similar to a system with
virtual memory \cite{virtualmemory}.

\begin{figure}
\includegraphics[width=11cm]{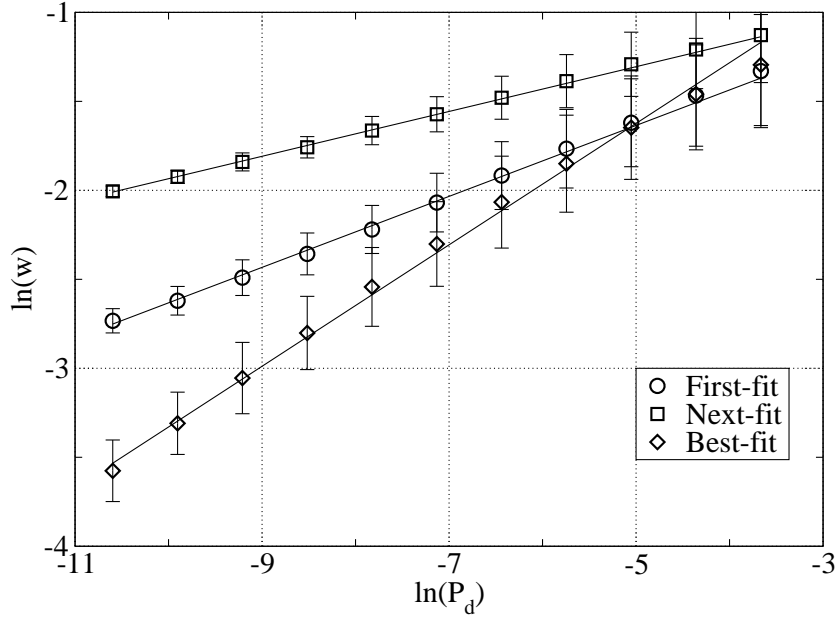}
\caption{\label{wasteline} Logarithm of wasted memory Vs $ln(P_{d})$ for
$S_{max}=256$ and three method: first-fit, next-fit and best-fit.}
\end{figure}
\begin{figure}
\includegraphics[width=11cm]{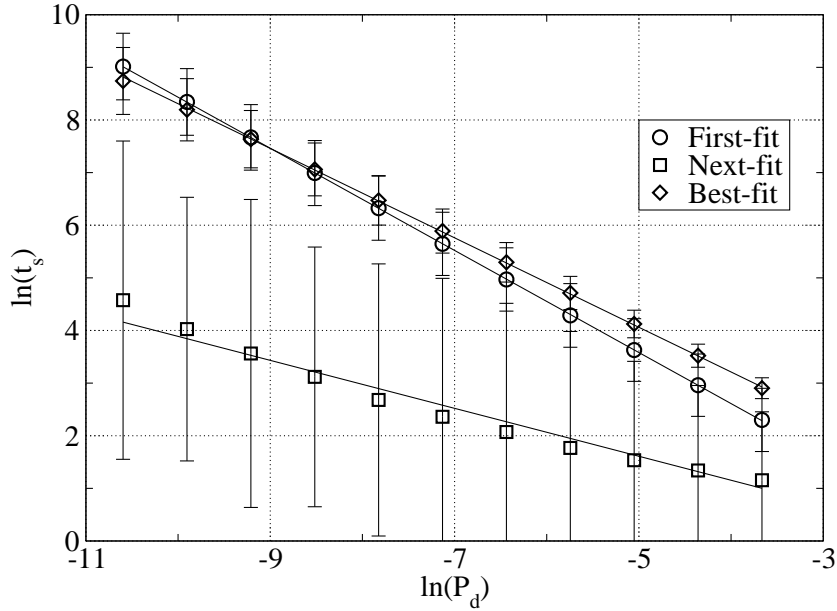}
\caption{\label{timeline} Logarithm of searching time Vs $ln(P_{d})$ for $S_{max}=256$. }
\end{figure}

The simulation runs until the system reaches the stationary state. Then one starts to measure the density of
allocated memory on the device and the searching time.

Density, $d$, is defined as the ratio of the total amount of allocated memory to the size of our flexible
storage device, including empty spaces; therefore wasted space, $w$, is simply $w=1-d$.

The searching time, $t_s$, is the time that allocator spends to find a free block and allocate requested memory.
The unit of time is the time needed to find a free block and check its size.
Thus allocation time is the number of checks which have to be done through the free list to find the
appropriate free block which satisfies all dictated constrains by the algorithm.
For example, in the first-fit method, if the first block in the list is satisfactory, this value will be
one, if the second block is, it will be 2 and so on. The conversion of this time unit to real time
depends on the hardware and many other characteristics of the system and it could be
studied as an independent problem for particular uses.

Averages of the two quantities $w$ and $t_s$ are calculated and stored
for all three algorithms. All six cases are performed for different value of $P_d$ and $S_{max}$.

\section{Results}
In the case of wasted memory a scaling ansatz which works rather well in all three methods is
\begin{equation}
w = P_d^{\nu_m} f_m(S_{max} P_d^{\phi_m}),
\end{equation}
with $f_m(x)$ being a constant for large value of $x$.
Similarly to the same form,
\begin{equation}
t_s= P_d^{\nu_t} f_t(S_{max} P_d^{\phi_t}),
\end{equation}
fits the data of searching time.

To have an estimate of the exponents $\nu_m$ we ran the simulation for a fix value of $S_{max}$
and various $P_d$ with $x=S_{max} P_d^{\phi_m}$ being large enough to be sure that $f_m(x)$ is constant.
The results are plotted in figure \ref{wasteline}, which shows the relation between wasted space and $P_d$
for three algorithms. The exponents, $\nu_m$, are the
slopes of the lines.

The same method has been applied to the searching times and results are shown in figure \ref{timeline}.
The value of exponents $\nu_m$  and $\nu_t$ are summarized in table \ref{exponents}.

Once the values of $\nu_m$ and $\nu_t$ are known, the functions $f_m(S_{max} P_d^{\phi_m})$
and $f_t(S_{max} P_d^{\phi_t})$ can be calculated to get the exponents $\phi_m$ and $\phi_t$.
We repeat the simulation for different $S_{max}$ for all the algorithms and at least three different values of $P_d$.

\begin{figure}
\includegraphics[width=12cm]{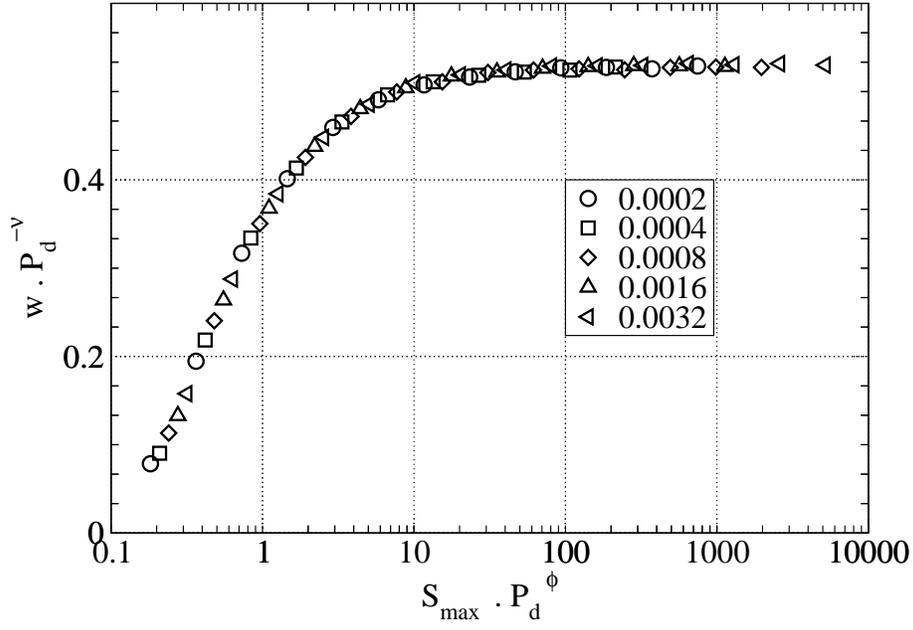}
\caption{\label{next_scaled} Scaled curves of wasted memory for next-fit method. Each symbol is for one $P_d$.}
\end{figure}

\begin{figure}
\includegraphics[width=15cm]{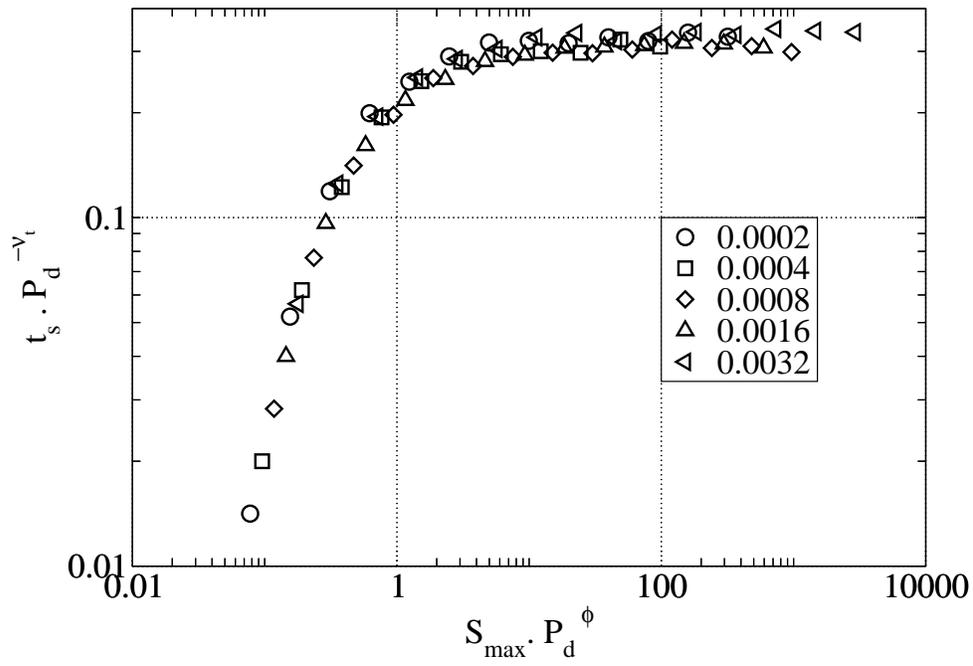}
\caption{\label{next_speed_scaled} Scaled curves of searching time for next-fit method. Each symbol is for one $P_d$. }
\end{figure}

For a given algorithm and with a properly selected value for the $\phi$ exponent, all the curves collapse.
Figure \ref{next_scaled} shows the collapse  for wasted space and figure \ref{next_speed_scaled}
for the searching time  just for the next-fit algorithms. Similar results hold for the first and best fit.
All of these six sets of data collapse fairly well with their own scaling functions and with the exponents
reported in table \ref{exponents}.

\begin{table}

\begin{tabular}{|l|c|c|c|c|}
\hline
        &\multicolumn{2}{c|}{$w$} &\multicolumn{2}{c|}{$t_s$} \\
\hline
Algorithm & $\nu_m$ & $\phi_m$ & $\nu_t$ & $\phi_t$ \\
\hline
First-fit & 0.21 & 0.25 & -1 & 0.23 \\
\hline
Next-fit &  0.13& 0.20 &  -0.5 & 0.3 \\
\hline
Best-fit &  0.35 & 0 & -1 & 0.5 \\
\hline
\end{tabular}
\caption{\label{exponents}Exponents related to all three algorithms for both cases of space usage and running speed.
The errors are at most $\pm 0.05$ }
\end{table}

Interestingly we note that, for the case of best-fit algorithm we obtain two values for $\phi$ whose difference is
substantially larger than the errors if we try to collapse
wasted space data or searching time data. In critical phenomenon theory, one expects that a unique
exponent is associated to each scaling field. However, in the present context we have no explanation for the results
that there may be more than one exponent associated with $P_d$.

Figure \ref{wasteline} proves that the best-fit algorithm can save more memory when the number of allocated blocks, $N_b$, is high (or
$P_d$ is small). For a smaller number of $N_b$, all three algorithms are similar. An important point
about best-fit algorithm is that it does not have any characteristic size because $\phi_m=0$ for this case.

On the other hand, for the searching time (FIG.\ref{timeline}) the next fit algorithm performs better than
the other two, which show a very similar behavior to one another. An interesting point is that the searching time of
next-fit grows with root square of $N_b$, while the searching time of other two grows linearly.

\section{Conclusion}
These examples can show us how the knowledge of the
scaling function and critical exponents of algorithms help us to choose an optimal strategy. Therefore,
the illustrated work can be repeated for other storage allocation algorithms
and different distributions of requests' sizes which
may be taken out from actual statistics of a particular system.
Nonetheless, it might be useful to apply the same method to
other complex algorithms, especially in the case that an algorithm should be
selected or designed for a large scale system like an Internet search engine.
In addition, we can study algorithms from this point of view
and classify them and determine whether any universality exists.
Finally one might extend this to other applied fields and engineerings
to organize data in more efficient way.

\section{Acknowledgment}
I thank Amos Maritan for valuable advice and a careful reading of
the manuscript.I thank Muli Safra for his comments.
\bibliographystyle{elsart-num}
\bibliography{computer}
\end{document}